\documentclass{article}

\usepackage[T1]{fontenc}
\usepackage{lmodern}

\usepackage[english]{babel}
\usepackage{csquotes}
\usepackage[babel]{microtype}

\usepackage[
backend=biber,
style=alphabetic,
sorting=ynt
]{biblatex}

\usepackage{amsmath,amssymb,amsthm}
\usepackage{mathtools}
\usepackage{xcolor}
\usepackage{tikz-cd,tikz}
\usepackage{tensor}
\usepackage{hyperref}
\usepackage[nameinlink]{cleveref}
\usepackage{todonotes}
\usepackage[top=3.33cm, left=3.33cm, right=3.33cm, bottom=3.33cm]{geometry}
\usepackage{enumitem}

\makeatletter
\mathchardef\ordinarycolon\mathcode`\:%
\mathcode`\:=\string"8000%
\begingroup \catcode`\:=\active%
	\gdef:{\@ifnextchar{=}
		{\coloneq\@gobble}
		{\ordinarycolon}}%
\endgroup
\AtBeginDocument{\mathcode`\==\string"8000} 
\mathchardef\ordinaryequals\mathcode`\=%
\begingroup \catcode`\==\active%
	\gdef={\@ifnextchar{:}
		{\eqcolon\@gobble}
		{\ordinaryequals}
	}%
\endgroup
\makeatother

\theoremstyle{plain}
\newtheorem{thm}{Theorem}[section]
\newtheorem{lem}[thm]{Lemma}
\newtheorem{prop}[thm]{Proposition}

\theoremstyle{definition}
\newtheorem{defn}[thm]{Definition}
\newtheorem{cons}[thm]{Construction}

\newtheorem{nota}[thm]{Notation}
\newtheorem{exmp}[thm]{Example}

\newtheorem{setup}[thm]{}
\theoremstyle{remark}
\newtheorem{rem}[thm]{Remark}

\crefname{lem}{lemma}{lemmas}

\AddToHook{env/lem/begin}{\crefalias{thm}{lem}}
\AddToHook{env/prop/begin}{\crefalias{thm}{prop}}
\AddToHook{env/obs/begin}{\crefalias{thm}{obs}}
\AddToHook{env/cor/begin}{\crefalias{thm}{cor}}
\AddToHook{env/conj/begin}{\crefalias{thm}{conj}}
\AddToHook{env/defn/begin}{\crefalias{thm}{defn}}
\AddToHook{env/cons/begin}{\crefalias{thm}{cons}}
\AddToHook{env/con/begin}{\crefalias{thm}{con}}
\AddToHook{env/ter/begin}{\crefalias{thm}{ter}}
\AddToHook{env/ass/begin}{\crefalias{thm}{ass}}
\AddToHook{env/nota/begin}{\crefalias{thm}{nota}}
\AddToHook{env/exmp/begin}{\crefalias{thm}{exmp}}
\AddToHook{env/fact/begin}{\crefalias{thm}{fact}}
\AddToHook{env/sol/begin}{\crefalias{thm}{sol}}
\AddToHook{env/setup/begin}{\crefalias{thm}{setup}}
\AddToHook{env/rem/begin}{\crefalias{thm}{rem}}
\AddToHook{env/note/begin}{\crefalias{thm}{note}}
\AddToHook{env/pro/begin}{\crefalias{thm}{pro}}

\numberwithin{equation}{section}

\DeclareMathOperator{\id}{id}

\newcommand{\ev}{\mathrm{ev}}

\newcommand{\R}{\mathbb R}

\newcommand{\C}{\mathbb C}
\newcommand{\I}{\mathrm i}
\newcommand{\N}{\mathbb N}
\newcommand{\g}{\mathfrak g}
\newcommand{\h}{\mathfrak h}
\newcommand{\n}{\mathfrak n}

\DeclareMathOperator{\im}{im}

\DeclareMathOperator{\GL}{GL}
\newcommand{\Lin}{\mathcal L}
\newcommand{\Lf}{\mathbf L}
\newcommand{\enter}{\vspace{\baselineskip}}
\newcommand{\henter}{\vspace{0.5\baselineskip}}

\title{Group Contractions via Infinite-Dimensional Lie Theory}
\author{David Prinz\footnote{Max Planck Institute for Mathematics, Bonn; e-mail: \href{mailto:prinz@mpim-bonn.mpg.de}{prinz@mpim-bonn.mpg.de}}, Alexander Schmeding\footnote{Norwegian University of Science and Technology, Trondheim; e-mail: \href{mailto:alexander.schmeding@ntnu.no}{alexander.schmeding@ntnu.no}} \kern0.2em and Philip K. Schwartz\footnote{Leibniz University Hannover; e-mail: \href{mailto:philip.schwartz@itp.uni-hannover.de}{philip.schwartz@itp.uni-hannover.de}}}
\date{June 20, 2026}

\begin{document}

\maketitle

\begin{abstract}
	Contractions are a procedure to construct a new Lie algebra out of a given one via a singular limit. Specifically, the \.{I}nönü--Wigner construction starts with a Lie algebra \(\mathfrak{g}\) with Lie subalgebra \(\mathfrak{h} \subseteq \mathfrak{g}\) and complement \(\mathfrak{n}\). Then, the vectors in \(\mathfrak{n}\) are rescaled by a formal parameter \(\varepsilon \in \mathbb{R}_+\), which effectively turns the Lie bracket \([ \, \cdot \, , \cdot \, ]\) into an \(\varepsilon\)-dependent family \([ \, \cdot \, , \cdot \, ]_\varepsilon\). Notably, the limit \(\varepsilon \to 0\) trivializes certain relations, such that the complement \(\mathfrak{n}\) becomes an abelian ideal. In the present article, we are not only interested in the limiting Lie algebras and groups, but also in the corresponding power series expansions in \(\varepsilon\) to understand their limiting behavior. Particularly, we are interested in the integration of the `power-series-expanded' Lie algebras to their corresponding Lie groups. To this end, we reformulate the above procedure using infinite-dimensional Lie algebras of analytic germs and then apply their integration theory. Our main results are a construction of the corresponding `Lie group expansions' in terms of quotients of groups of analytic germs and an explicit description of these groups in elementary terms. Applications of this procedure include the geometric Newtonian limit of General Relativity to Newton--Cartan gravity, where the Poincaré group is contracted to the Galilei group.
\end{abstract}

\henter

\smallskip


\tableofcontents

\section{Introduction} \label{sec:introduction}

The general idea of Lie algebra contractions goes back to Segal's work \cite{Seg51} and was then studied in a more specialized way by Wigner and \.{I}nönü \cite{Inonu:1953sp}: A Lie algebra contraction aims to
construct a new Lie algebra from a given one via a \emph{singular limit} procedure. Let us first exemplify this by the standard example, the contraction of \(\mathfrak{so}(3)\) to \(\mathfrak{iso}(2)\). We present this example with explicit
coordinates and then turn to a coordinate-free notation later. Consider \(\mathfrak{so}(3)\) with generators \(X_1, X_2,
X_3\). Then, the structure constants for \(\mathfrak{so}(3)\) are given by the Levi-Civita symbol in three dimensions, i.e.\
\(\tensor{f}{_{ab}^c} = \tensor{\epsilon}{_{ab}^c}\) or, explicitly,
\begin{equation}
	[X_1, X_2] = X_3 \, , \quad [X_2, X_3] = X_1 \quad \text{and} \quad [X_3, X_1] = X_2 \, .
\end{equation}
Now, we consider the rescalings
\begin{equation}
	Y_1 := \varepsilon X_1 \, , \quad Y_2 := \varepsilon X_2 \quad \text{and} \quad Y_3 := X_3 \, .
\end{equation}
This results in the Lie bracket relations
\begin{equation}
	[Y_1, Y_2] = \varepsilon^2 Y_3 \, , \quad [Y_2, Y_3] = Y_1 \quad \text{and} \quad [Y_3, Y_1] = Y_2 \, .
\end{equation}
Notably, the limit \(\varepsilon \to 0\) trivializes the first commutator and produces the Lie algebra \(\mathfrak{iso}(2)\), the Lie algebra of the Euclidean group in two dimensions. Using the results established in the present article, we discuss the corresponding group level in \Cref{exmp:SO3toISO2}.

\enter

Generally, the notion of \.{I}nönü--Wigner Lie algebra contraction (in the literature often misleadingly referred to as \emph{group contraction}) is defined as follows. Consider a finite-dimensional\footnote{For the present introduction, we restrict to the finite-dimensional case, while allowing infinite-dimensional Lie algebras later in the article.} Lie algebra \(\g\) with a Lie subalgebra \(\h \subseteq \g\), and any vector space complement \(\n\), i.e.
\begin{equation}
	\g = \h \oplus \n \quad \text{as a vector space.}
\end{equation}
We then define a one-parameter family of linear endomorphisms \(\phi_\varepsilon \colon \mathfrak{g} \to \mathfrak{g}\), \(\varepsilon \in [0,1)\), by
\begin{equation} \label{eq:iw-contraction}
	\phi_\varepsilon (X_\h + X_\n) := X_\h + \varepsilon X_\n \, ,
\end{equation}
where \(X = X_\h + X_\n\) is the decomposition of a general element \(X \in \mathfrak{g}\) corresponding to the decomposition $\g = \h \oplus \n$ with \(X_\h \in \h\) and \(X_\n \in \n\). The contraction of \(\g\) with respect to \(\h\) is then \(\g\) equipped with the Lie bracket that arises as the singular \(\varepsilon \to 0\) limit of
\begin{equation}
	[X,Y]_\varepsilon := \phi_\varepsilon^{-1} ([\phi_\varepsilon(X), \phi_\varepsilon(Y)]_\g) \, .
\end{equation}
In particular, this limit \(\varepsilon \to 0\) turns \(\mathfrak{n}\) into an abelian ideal. More generally than this case discussed by \.{I}nönü and Wigner, one may consider arbitrary families $\phi_\varepsilon$ of linear endomorphisms that are invertible for $\varepsilon > 0$.

\enter

Lie algebra contractions and the related concept of Lie algebra deformations \cite{CM_1990__75_1_69_0,Bur07} are commonly studied in terms of the $\varepsilon$-dependent Lie bracket family $[ \, \cdot \, , \cdot \, ]_\varepsilon$. In this context, the asymptotic behavior of the singular $\varepsilon \to 0$ limit has been studied e.g.\ in \cite{Khasanov.Kuperstein:2011}: here, one understands $(\g, [ \, \cdot \, , \cdot \, ]_\varepsilon)$ as a bundle of Lie algebras over $[0,1)$, considers sections in this bundle, and uses these to study the limit $\varepsilon \to 0$ by power series expansions in $\varepsilon$. Concretely, the space of sections is an infinite-dimensional extension of $\g$, quotients of which encode the asymptotic behavior of the contraction.

The associated group level is often not directly considered. To our knowledge, the most thorough study of contractions of groups uses group schemes and algebraic groups in \cite{Dab99}. There it was shown that the Lie algebra contractions induce a corresponding operation for algebraic groups. These results use Hopf algebraic methods to describe the convergence on the group level. Note that this approach is purely algebraic as analytic concerns are fixed by the algebraic group structure.

\enter

Our goal in the present article is to understand the \emph{global} picture, i.e.\ lift the above construction of `Lie algebra expansions' \cite{Khasanov.Kuperstein:2011} from the Lie \emph{algebra} to the Lie \emph{group} level. For this purpose, we will use the language of infinite-dimensional Lie theory. An equivalent perspective to that of \cite{Khasanov.Kuperstein:2011} turns out to be more suitable: We still consider $\varepsilon$-dependent curves in $\g$, but to these we apply only the original Lie bracket $[ \, \cdot \, , \cdot \, ]_\g$. The family of `rescaling maps' $\phi_\varepsilon$ comes into play regarding \emph{which} curves in $\g$ we consider.

To this end, we consider the Lie algebra $\Gamma(\g)$ of germs of analytic curves in $\g$. By $\Gamma(\g)_\phi \subseteq \Gamma(\g)$, we denote the subset consisting of germs that arise by `pointwise application of $\phi$', i.e.
\begin{equation}
	\Gamma(\g)_\phi := \{[c] \in \Gamma(\g) \mid \exists [\tilde{c}] \in \Gamma(\g) : c(\varepsilon) = \phi_\varepsilon(\tilde{c}(\varepsilon)) \} \, .
\end{equation}
This may be shown to be a subalgebra of $\Gamma(\g)$, cf.\ \Cref{defn:Lie_alg_contr_germ} (i). In this subalgebra, the sets $\varepsilon^k \Gamma(\g)_\phi$ are ideals. The quotients
\begin{equation}
	\g_\phi^{(k)} := \Gamma(\g)_\phi \Big/ \varepsilon^{k+1} \Gamma(\g)_\phi \, ,
\end{equation}
which we call \emph{Lie algebra expansions} of order $k$, then encode the asymptotic behavior of the contraction: the order-$0$ expansion $\g_\phi^{(0)}$ is naturally isomorphic to the contraction $\g_0$, and the higher-order quotients encode \emph{how} the limit is approached in a power-series expansion. In the case of \(\phi\) being the
\.{I}nönü--Wigner map from \Cref{eq:iw-contraction}, we also simply write \(\Gamma(\g)_\h\) for the corresponding subalgebra of \(\Gamma(\g)\), and likewise \(\g_\h^{(k)}\) for the corresponding expansions.

Note that our approach is philosophically close to the van Est construction for the integration of a Lie algebra to a Lie group, cf.\ e.g.\ \cite{DK00}. There, certain loops with values in a Lie algebra are considered and a quotient is formed to obtain the integrating finite-dimensional Lie group. However, in our approach, we use germs of analytic functions. Our methods are weaker than the van Est construction, but we also do not need cohomological data for the construction.

\enter

\enlargethispage{1.5\baselineskip}

Our first main result is then a lift to the Lie group level, for the \.{I}nönü--Wigner special case:
\begin{thm}[Informal version of first main result (\Cref{thm:main_result})]
	Let $G$ be a (real) Banach Lie group with Lie algebra $\g = \Lf(G)$, and let $H \subseteq G$ be a Lie subgroup with Lie algebra $\h = \Lf(H)$. Denote by $\Gamma(G)$ the BCH Lie group of germs of analytic curves in $G$.
	\begin{enumerate}[label=(\roman*)]
		\item There is an analytic Lie subgroup
		\begin{equation}
			\Gamma(G)_H \subseteq \Gamma(G)
		\end{equation}
		of $\Gamma(G)$ with Lie algebra $\Lf(\Gamma(G)_H) = \Gamma(\g)_\h$.
		\item For any $k \in \N_0$, there is a normal analytic Lie subgroup
		\begin{equation}
			\Gamma(G)_H^{(k+1)} \trianglelefteq \Gamma(G)_H
		\end{equation}
		of $\Gamma(G)_H$ with Lie algebra $\Lf(\Gamma(G)_H^{(k+1)}) = \varepsilon^{k+1} \Gamma(\g)_\h$.
		
		Hence, the quotient
		\begin{equation}
			G^{(k)}_H := \Gamma(G)_H \Big/ \Gamma(G)_H^{(k+1)}
		\end{equation}
		is a BCH Lie group whose Lie algebra is the order-$k$ Lie algebra expansion, $\Lf(G^{(k)}_H) = \g^{(k)}_\h$.
	\end{enumerate}
\end{thm}
Our second main result (\Cref{thm:k-th-expansion}) explicitly constructs the Lie group expansions $G^{(k)}_H$ as Banach Lie groups.

\enter

\paragraph{Applications in mathematics:}
The present construction allows for various new perspectives on symmetries, their deformations, and contractions within mathematics. Notably, expansions of order \(k + 1\) can be seen as deformations of the expansions of order \(k\).

In addition, the Lie algebra expansions \(\g_\phi^{(k)}\) inherit the structure of a \emph{nilpotent graded} Lie algebra: this can be achieved by considering the parameter \(\varepsilon\) as an element of the graded line, i.e.\ \(\varepsilon \in \R[1]\), inducing a natural gradation on the germ Lie algebra \(\Gamma(\g)\) by power series expansion of germs. Concretely, in the \.{I}nönü--Wigner case \eqref{eq:iw-contraction} the induced gradation on the order-\(k\) Lie algebra expansion reads
\begin{subequations}
\begin{align}
	\g_\h^{(k)} & \cong \h \oplus \left( \bigoplus_{\ell=1}^k \varepsilon^\ell \g \right) \oplus \varepsilon^{k+1} (\g/\h)
	\intertext{(with higher grades vanishing). In particular, the order-\(0\) expansion (i.e., up to natural isomorphism, the contraction $\g_0$) takes the form}
	\g_\h^{(0)}
	& \cong \h \oplus \varepsilon \, (\g/\h) \, .
\end{align}
\end{subequations}
Note that due to \(\varepsilon \, (\g/\h)\) forming an abelian ideal, we may view \(\g_\h^{(0)}\) also as a (rather trivial) \emph{super} Lie algebra, viewing \(\varepsilon\) as an element of the odd line, \(\varepsilon \in \Pi \R\). In particular, these identifications allow us to apply the integration theory for \emph{super} or \emph{nilpotent graded} Lie algebras to their respective \emph{super} or \emph{nilpotent graded} Lie groups using \emph{Harish-Chandra modules}, cf.\ \cite{SCHEUNERT2005324,Gavarini}. This perspective will be studied in a follow-up article. We believe that this formulation, using a different language, provides a valuable addition to our present results.

Further, we plan to construct a \emph{derived} version, which measures the respective deformations explicitly using the dual perspective of the Chevalley--Eilenberg cochain complex, cf.\ \cite{Jubin_Kotov_Poncin_Salnikov}. More generally, we plan to generalize the present constructions to Lie algebroids and Lie groupoids as well as \(L_\infty\)-algebras and \(L_\infty\)-groups, cf.\ \cite{Getzler}.

\paragraph{Applications in physics:}
The original motivation for Lie algebra contractions comes from physics. Here, contractions play a role in any situation where two physical theories are related by a singular limit which affects in particular symmetry transformations. Most famously, this regards the Newtonian limits of General and Special Relativity to Newtonian physics: in the `relativistic' case, the relevant groups are the Lorentz and/or Poincaré (i.e.\ inhomogeneous Lorentz) groups, which the singular limit $1/c \to 0$, where $c$ is the speed of light, contracts to the homogeneous and inhomogeneous Galilei groups. This contraction `Poincaré $\leadsto$ Galilei' was the original case considered by \.{I}nönü and Wigner \cite{Inonu:1953sp}. More generally, the classification of possible kinematical groups (i.e.\ groups that are allowed as spacetime symmetries) by Bacry and Lévy-Leblond \cite{Bacry.Levy-Leblond:1968} has shown that all permitted kinematical groups arise via a sequence of contractions of the de~Sitter and Anti-de~Sitter groups.

A fully coordinate-free geometric description of the `Newtonian $1/c \to 0$ limit' of General Relativity and Lorentzian geometry is given by so-called \emph{Newton--Cartan gravity} and \emph{Galilei geometry} \cite{Cartan:1923,Cartan:1924,Trautman:1963,Kuenzle:1972,Kuenzle:1976,Ehlers:1981a,Ehlers:1981b,Schwartz:NC_gravity}. In the limit, the bundle of Lorentzian orthonormal frames is replaced by the bundle of so-called Galilei frames, whose structure group is the homogeneous Galilei group---i.e., heuristically speaking, the contraction happens at the level of structure groups of bundles of adapted frames. In this context, the notion of Lie algebra \emph{expansions} \cite{Khasanov.Kuperstein:2011} has recently been used to give a coordinate-free description of the post-Newtonian expansion of General Relativity in a post-Newton--Cartan formalism, by considering connections valued in Lie algebra expansions of the Lorentz / Poincaré algebra corresponding to their contractions to the Galilei algebras \cite{Hansen.EtAl:2019,Hansen.EtAl:2020,Hartong.EtAl:2023}. The results of the present paper open up the interesting possibility to extend this description rigorously to the level of Lie \emph{groups}.

Further, it would be interesting to apply our results to the Lie groups arising as asymptotic symmetry groups of asymptotically flat spacetimes, such as the \(\operatorname{BMS}(3,1)\) and \(\operatorname{NU}(3,1)\) groups, cf.\ \cite{Prinz:2021bnw,Prinz:2021afr}. However, as the BMS and NU groups are semidirect products of a finite-dimensional Lie group and a space of smooth functions, they are not Banach Lie groups. Even more severe, these groups are not locally exponential Lie groups (in the sense that their Lie group exponential does not provide exponential coordinates around the identity) \cite{CaDaNaS25}, whence our methods are not applicable. However, as the space of smooth functions is a tame Fréchet space \cite{Ham82}, the BMS group is a benign inverse limit of Banach spaces. Conceivably, the ideas developed in this article together with methods for inverse limit groups (à~la Omori's treatment of the diffeomorphism group \cite{Omo74}) will allow one to treat contractions and expansions thereof of at least the BMS group. We leave this to future work.

\paragraph{This article is structured as follows:}
We start in \Cref{sec:preliminaries-general-setup} with general Lie algebra contractions and the necessary background on infinite-dimensional Lie theory: Especially, this includes Lie algebras and Lie groups of germs of real-analytic mappings. Then, we proceed in \Cref{sec:LAC-germ-algebras} by reformulating Lie algebra contractions and expansions in terms of germ algebras: Notably, in \Cref{prop:Lie_alg_expansion_0_contr} we identify the contracted Lie algebra with the zero-order expansion defined via analytic germs. Furthermore, in \Cref{rem:higher-expansions-equivalence}, we discuss the meaning of the corresponding higher-order expansions. Next, in \Cref{sec:LGC-germ-groups}, we study the corresponding Lie group level: Specifically, this includes our first main result \Cref{thm:main_result}, which lifts all constructions performed previously on the level of germ Lie algebras to the corresponding germ Lie groups. Moreover, in \Cref{thm:k-th-expansion}, we use this to explicitly identify the Lie group expansions as specific Banach Lie groups. Finally, in \Cref{sec:conclusion}, we conclude our findings and give an outlook on planned follow-up projects.

\section{Preliminaries and general setup} \label{sec:preliminaries-general-setup}

In this section, we lay down our basic notations and conventions. Furthermore, we review the basics of Lie algebra contractions, as well as some essentials on groups and algebras of germs of analytic functions. These form infinite-dimensional Lie groups in the sense of \cite{Neeb06,Sch23}.

\begin{nota}
	In general, we write $\N = \{1,2,3,\ldots\}$ for the set of natural numbers and $\N_0 = \N\cup \{0\}$ if we start from zero. Unless otherwise stated, all vector spaces and manifolds are over the field $\R$ of real numbers. Differentiability is understood in the sense of Bastiani calculus (see, e.g., \cite{Sch23} for an introduction) unless otherwise stated.
\end{nota}

We present the theory first as generally as possible before specializing to more concrete applications. In defining analyticity for mappings taking values in infinite-dimensional spaces we follow \cite{BaS71}:

\begin{defn}
	Let $E$ be a complex topological vector space, $F$ a complex locally convex space, and $U \subseteq E$ open and non-empty. The map $f\colon U \rightarrow F$ is called \emph{complex analytic} if it is continuous and for every $x \in U$, there exists a zero-neighborhood $V \subseteq E$ such that $x + V\subseteq U$ and
	\begin{equation}
		f(x+h) = \sum_{n=0}^\infty \beta_n (h) \quad \text{for all} \; h \in V \; \text{as pointwise limit},
	\end{equation}
	for suitable continuous homogeneous polynomials $\beta_n \colon E \rightarrow F$ of degree $n \in \N_0$.\footnote{Recall that a map $p \colon E \rightarrow \C$ is called a homogeneous polynomial of degree $n$ if there is a continuous $n$-linear map $q \colon E^n \rightarrow \C$ such that $p(x) = q(x, \dots, x)$.}
\end{defn}

\begin{defn}
	Let $E$ be a real topological vector space, $F$ a real locally convex space, and $U \subseteq E$ open and non-empty. A map $f\colon U \rightarrow F$ is called \emph{(real) analytic} if $f$ extends to a complex-analytic map $V \rightarrow F_\C$ on some open neighborhood $V$ of $U$ in the complexification $E_\C = E \oplus \I E$ of $E$.
\end{defn}

\subsection{Lie algebra contractions}

The \emph{general} notion of Lie algebra contractions (cf., e.g., the discussions in \cite{Dab99} and \cite{Bur07}) is as follows.
\begin{defn}
	Let $\g$ be a locally convex Lie algebra.\footnote{A Lie algebra is locally convex if it is also a locally convex (Hausdorff) space such that the Lie bracket is continuous, cf.\ \cite{Neeb06,Sch23}. Every finite-dimensional Lie algebra is automatically locally convex.} Let
	\begin{equation}
		\phi \colon [0,a) \to \Lin(\g,\g), \; \text{such that for} \; \varepsilon \neq 0, \; \phi_\varepsilon := \phi(\varepsilon) \in \GL(\g)
	\end{equation}
	be a map into the continuous linear endomorphisms of $\g$. ($\GL(\g)$ denotes the group of continuous linear automorphisms of $\g$ with continuous inverse.) This defines a one-parameter family of Lie brackets on $\g$, depending on $\varepsilon \ne 0$, by
	\begin{equation} \label{eq:contr_bracket_rescale}
		[X,Y]_\varepsilon := \phi_\varepsilon^{-1} ([\phi_\varepsilon(X), \phi_\varepsilon(Y)]_\g)
	\end{equation}
	and we set \(\g_\varepsilon := (\g, [ \, \cdot \, , \cdot \, ]_\varepsilon)\). If the pointwise limit
	\begin{equation} \label{eq:contr_bracket_limit}
		[ \, \cdot \, , \cdot \, ]_0 := \lim_{\varepsilon \to 0} \, [ \, \cdot \, , \cdot \, ]_\varepsilon
	\end{equation}
	exists, i.e.\ the limit of $[ \, \cdot \, , \cdot \, ]_\varepsilon$ exists when evaluated on any two elements of $\g$, then this limit $[ \, \cdot \, , \cdot \, ]_0$ defines a new Lie bracket on $\g$. The corresponding Lie algebra $\g_0 := (\g, [ \, \cdot \, , \cdot \, ]_0)$ is the \emph{Lie algebra contraction of $\g$ with respect to $\phi$}.
\end{defn}

\begin{rem}
	\begin{enumerate}[label=(\roman*)]
		\item For each $\varepsilon \ne 0$, by definition we have a locally convex Lie algebra isomorphism (i.e.\ a continuous Lie algebra homomorphism with continuous inverse) $\phi_\varepsilon \colon \g_\varepsilon \xrightarrow{\cong} \g$.
		\item By taking the $\varepsilon \to 0$ limit of \eqref{eq:contr_bracket_rescale}, we obtain that $\phi_0 \colon \g_0 \to \g$ is a Lie algebra homomorphism. The interesting case of contractions are the non-trivial ones, for which $\phi_0 \in \Lin(\g,\g)$ is \emph{not} invertible.
		
		The contraction $\g_0$ is an extension of the subalgebra $\h := \im\phi_0 \subseteq \g$ by the ideal $\n := \ker\phi_0 \trianglelefteq \g_0$, i.e.\ we have the following short exact sequence:
			\begin{equation}
			\begin{tikzcd}
				0 \arrow[r] & \n \arrow[r] & \g_0 \arrow[r,"\phi_0"] & \h \arrow[r] & 0
			\end{tikzcd}
			\end{equation}
			In the finite-dimensional case, using Engel's theorem one may show that $\n$ is nilpotent \cite[Theorem 1.28]{Dab99}.
		\item The original case of contractions as discussed by \.{I}nönü and Wigner is as follows.
		
		Starting with a subalgebra $\h \subseteq \g$ of the original algebra and a vector space complement $\n \subseteq \g$, such that
			\begin{equation}
				\g = \h \oplus \n \; \text{as a vector space},
			\end{equation}
			we define $\phi_\varepsilon \colon \g \to \g$ by
			\begin{equation}
				\phi_\varepsilon(X_\h + X_\n) := X_\h + \varepsilon X_\n
			\end{equation}
			for $X_\h \in \h, X_\n \in \n$. Then the contraction $\g_0$ of $\g$ with respect to $\phi$ exists, and in $\g_0$ the complement $\n$ is an \emph{abelian} ideal.

		One may show that this contraction is independent (up to isomorphism) of the choice of complement $\n$. In particular, this will follow from our discussion in \Cref{prop:Lie_alg_expansion_0_contr} and \Cref{cons:Inonu--Wigner} below.
	\end{enumerate}
\end{rem}

To apply the machinery of infinite-dimensional Lie theory, we need to restrict to the case of \emph{Banach} Lie groups, and consider \emph{analytic} contractions, in the following sense:
\begin{defn} \label{defn:contr_analytic}
	Let $\g$ be a Banach Lie algebra. Let $\phi \colon (-a,a) \to \Lin(\g,\g), \varepsilon \mapsto \phi_\varepsilon$ be a continuous one-parameter family\footnote{Note that different to before, here the parameter $\varepsilon$ takes values in an \emph{open} interval $(-a,a)$ around $0$, as compared to the half-open $[0,a)$ from before. This is in order to allow us to demand analyticity of $\varepsilon \mapsto \phi_\varepsilon$, particularly in $0$.} of (continuous) linear endomorphisms of $\g$ such that for $\varepsilon \ne 0$ we have $\phi_\varepsilon \in \GL(\g)$. (Here continuity of $\phi$ is with respect to the operator norm on $\Lin(\g,\g)$.)

	Assume that the contraction of $\g$ with respect to $\phi$ exists. The contraction is called an \emph{analytic Lie algebra contraction} if the family $\varepsilon \mapsto \phi_\varepsilon$ is analytic, and the family $(-a,a) \ni \varepsilon \mapsto [ \, \cdot \, , \cdot \, ]_\varepsilon$ of Lie brackets on $\g$ (defined by \eqref{eq:contr_bracket_rescale} for $\varepsilon \ne 0$, and $[ \, \cdot \, , \cdot \, ]_0 := \lim_{\varepsilon \to 0} \, [ \, \cdot \, , \cdot \, ]_\varepsilon$) is analytic as well.
\end{defn}

\begin{rem}
	Note that if $\varepsilon \mapsto \phi_\varepsilon$ is analytic, then $\varepsilon \mapsto [ \, \cdot \, , \cdot \, ]_\varepsilon$ is automatically analytic for $\varepsilon \ne 0$---so the additional assumption is about \emph{analyticity at $0$}.
\end{rem}

\subsection{Elements of infinite-dimensional Lie theory}

Here, we recall general results on subgroups and quotients of locally exponential Lie groups (which may be infinite-dimensional) from \cite{Neeb06}. \emph{Locally exponential} Lie groups are Lie groups for which the Lie group exponential induces a local diffeomorphism around the identity. Even more restrictive, the class of \emph{Baker--Campbell--Hausdorff (BCH) Lie groups} requires that the BCH series converges locally in the Lie algebra and yields an analytic model of the group multiplication. For these Lie groups, many basic constructions from finite-dimensional Lie theory carry over to the infinite dimensional setting.

\begin{defn}[\cite{Neeb06}, Definition IV.3.2]
	Let $G$ be a locally exponential Lie group. A closed subgroup $H \subseteq G$ is a \emph{(locally exponential) Lie subgroup} of $G$ if it is a locally exponential Lie group with respect to the induced topology.\footnote{This definition makes sense since a topological group is a locally exponential Lie group in at most one way (which follows from the `automatic smoothness theorem', see \cite[Theorem IV.1.18 and Remark IV.1.22]{Neeb06}).}
\end{defn}
Note that Lie subgroups of BCH Lie groups are automatically analytic subgroups (i.e.\ BCH Lie groups such that the inclusion map is analytic), since any continuous homomorphism of BCH Lie groups is analytic (the BCH case of the `automatic smoothness theorem', see \cite[Proposition~2.4]{Glo02} and \cite[Theorem IV.1.18]{Neeb06}).

Differently to the finite-dimensional case, closed subgroups of locally exponential Lie groups are \emph{not} automatically Lie subgroups. However, for closed groups the various notions of `candidate Lie algebras' for these groups coincide:
\begin{defn}[\cite{Neeb06}, Proposition II.6.3 and Lemma IV.3.1]
	Let $G$ be a locally exponential Lie group. For any subgroup $H \subseteq G$ we define the set
	\begin{equation}
		\Lf^e(H) := \{X \in \Lf(G) \mid \exp_G(\R X) \subseteq H\} \, .
	\end{equation}
	If $H \subseteq G$ is closed, then $\Lf^e(H) \subseteq \Lf(G)$ is closed, and it coincides with
	\begin{equation}
		\Lf^d(H) := \{\alpha'(0) \mid \alpha \in C^1([0,1],G), \alpha(0) = \mathbf{1}_G, \im(\alpha) \subseteq H\} \, ,
	\end{equation}
	which is a Lie subalgebra of $\Lf(G)$.

	If $H \subseteq G$ is a Lie subgroup, then $\Lf^e(H) = \Lf^d(H) = \Lf(H)$ is its Lie algebra.
\end{defn}

Furthermore, these `candidate Lie algebras' may be used to characterize those closed subgroups of a locally exponential Lie group $G$ which are, in fact, Lie subgroups. Concretely, such a characterization is provided by \cite[Theorem IV.3.3]{Neeb06}, of which we will now establish a variant. In this theorem, condition \eqref{eq:subgroup_exp_coords} (see below) plays a role, which is formulated for an open zero-neighborhood $V \subseteq \Lf(G)$ such that $\exp_G(V)$ is open and ${\exp_G}|_V$ is a diffeomorphism onto its image. However, assuming $U \subseteq \Lf(G)$ is an open zero-neighborhood on which $\exp_G$ is injective such that $\exp_G(U)$ is open, then on $W := U \cap V$, we have that ${\exp_G}|_W$ is still a diffeomorphism onto an open set in $G$; and it is easy to check that if \eqref{eq:subgroup_exp_coords} holds for $U$ (instead of $V$), then it also holds for $W$. Thus a (somewhat more convenient to check) variant of \cite[Theorem IV.3.3]{Neeb06} is the following:
\begin{thm} \label{thm:Lie_subgroup}
	Let $G$ be a locally exponential Lie group. A closed subgroup $H \subseteq G$ is a Lie subgroup if and only if there exists an open zero-neighborhood $V \subseteq \Lf(G)$ such that $\exp_G(V) \subseteq G$ is open, ${\exp_G}|_V$ is injective, and
	\begin{equation} \label{eq:subgroup_exp_coords}
		\exp_G(V \cap \Lf^e(H)) = \exp_G(V) \cap H \, .
	\end{equation}
\end{thm}

\begin{rem}
	In the case of BCH Lie groups, \Cref{thm:Lie_subgroup} characterizes analytic subgroups.
\end{rem}

For BCH Lie groups, normal analytic subgroups yield well-behaved quotients \cite[Theorem 2.20 and Corollary 2.21]{Glo02}:
\begin{thm} \label{thm:quotient}
	Given a closed normal subgroup $N$ of a BCH Lie group $G$, the quotient $G/N$ is a BCH Lie group if and only if $N$ is a Lie subgroup of $G$.
\end{thm}

\subsection{Lie algebras and groups of germs}

Now we consider a special case of the Lie algebras and groups of germs of real analytic mappings discussed in \cite{Glo04,DGS14}, which we will use in the following.
\begin{defn}
	Let $\g$ be a Banach Lie algebra and $G$ a Banach Lie group. By $\Gamma(\g)$ we denote the (infinite-dimensional) Lie algebra of germs of analytic curves in $\g$ with the pointwise Lie bracket, i.e.
	\begin{equation}
		\Gamma(\g) := \{c \colon (-a,a) \to \g \; \text{analytic} \mid a \in \R\} / \sim
	\end{equation}
	where $\sim$ is equality after restriction to some neighborhood of $0$. Similarly, we denote by $\Gamma(G)$ the group (with respect to the pointwise group operations) of germs of $G$-valued real analytic curves.
\end{defn}

\begin{nota}
	We will mostly denote representatives of germs by $c,d$ (and variants of these letters) in the case of germs taking values in Lie algebras, and by $f,g$ (and variants of these letters) for germs taking values in Lie groups. Germs themselves will be denoted by writing a representative in square `equivalence class' brackets, e.g.\ $[c] \in \Gamma(\g)$.

	The parameter of a germ will be denoted by $\varepsilon$. We will also abuse notation and, for $k \in \N_0$, write $\varepsilon^k$ for the analytic function $(-a,a) \ni \varepsilon \mapsto \varepsilon^k \in \R$ or its germ. Then, given a germ $[c] \in \Gamma(\g)$, we obtain $\varepsilon^k [c] \in \Gamma(\g)$ as the germ of analytic maps $[\varepsilon \mapsto \varepsilon^k c(\varepsilon)]$.
\end{nota}
The space $\Gamma(\g)$---and, analogously, $\Gamma(G)$---carries a natural inductive limit topology with respect to a sequence of Banach spaces of holomorphic mappings on some (complex) neighborhood of $0 \in \R$. We mention in particular that the topology on $\Gamma(\g)$ is finer than the compact open topology, whence point evaluations are continuous. Details can be found in \cite[Appendix A]{DS15}.

\begin{setup}
	Before we continue, recall that a subspace $F$ of a locally convex space $E$ is called \emph{complemented} if there is a vector space complement, i.e.\ another subspace $H$ of $E$ such that $E \cong F \times H$ as locally convex spaces. It is well-known that this is equivalent to the existence of a continuous linear projector $\pi \colon E \rightarrow E$ such that $\pi(E) = F$ and $\pi \circ \pi = \pi$. As a consequence also the complement $H$ is complemented (with a corresponding complement given by $F$). If the complement $H$ is a Banach space in the subspace topology, we say that $F$ is \emph{co-Banach} (cf.\ \cite[Appendix A]{Sch23}). We remark that the co-Banach property is not symmetric (unless the ambient space $E$ is already a Banach space, but then all complemented subspaces are (co-)Banach).

	Similarly, we say that a Lie subalgebra $\h \subseteq \g$ of a locally convex Lie algebra $\g$ is a \emph{closed (or complemented, or co-Banach) Lie subalgebra} if it is closed (or complemented, or co-Banach) as a subspace in $\g$.
\end{setup}

We note the following consequence of \cite[A.10 and Lemma A.7]{DS15} for the space $\Gamma(\g)$:
\begin{lem} \label{Lem:ev}
	Let $\g$ be a Banach Lie algebra. Then point evaluation, the inclusion map, and the map sending a germ to the evaluation of its $k$-th derivative
	\begin{align}
		& \ev_0 \colon \Gamma(\g) \to \g, \quad && [c] \mapsto c(0), \\
	& C \colon \g \to \Gamma(\g), \quad && C(X) = [\varepsilon \mapsto X],\\
	& D^{(k)}_0 \colon \Gamma (\g) \to \g, \quad && [c] \mapsto c^{(k)}(0), \, k \in \N,
	\end{align}
	are continuous linear. The kernel of $\ev_0$ is complemented and the complement is isomorphic to $\g$ via $C$, whence the kernel is co-Banach.
\end{lem}

\begin{proof}
	Linearity of the maps $D^{(k)}_0$ is clear for every $k\in \N_0$ and with the convention $c^{(0)}=c$ for the $0$-th derivative we have $\ev_0=D^{(0)}_0$. Thus we shall prove continuity of $D^{(k)}_0, k\in \N_0$. The map $C$ makes sense and is clearly linear with $\ev_0 \circ C = \id_{\g}$. As cited, the topology on $\Gamma(\g)$ is induced by identifying $\Gamma(\g)$ with a closed real subspace of the locally convex inductive limit of a family of spaces of bounded holomorphic functions $\mathrm{BHol} (B^\C_{1/n}(0), \g_\C)$, $n \in \N$. Thus for the maps $D^{(k)}_0$ it suffices to establish continuity of the induced maps $\tilde{D}^{(k)}_0 \colon \mathrm{BHol} (B^\C_{1/n}(0), \g_\C) \rightarrow \g_\C, f \mapsto f^{(k)}(0)$, $n \in \N$ on the steps of the limit (which are Banach spaces with respect to the supremum norm). However, as the mapping sending a bounded holomorphic function to its $k$-th derivative in $\mathrm{BHol} (B^\C_{1/n}(0), \g_\C)$ is continuous due to the Cauchy estimates and the evaluation map is continuous in the supremum norm, continuity of the $\tilde{D}_0^{(k)}$- and thus also of the $D^{(k)}_0$-maps follows.

	As $C$ factors through the inclusion of $\mathrm{BHol}(B^\C_{1}(0), \g_\C)$, it suffices to prove that the map $\g \to \mathrm{BHol}(B^\C_{1}(0), \g_\C)$, $x \mapsto (z \mapsto x)$ is continuous. Now the space of bounded holomorphic mappings is a Banach space with respect to the supremum norm topology (cf.\ \cite[A.8]{DS15}), whence continuity of the mapping is clear. We deduce that $C$ is continuous linear and thus the kernel of $\ev_0$ is complemented with $\im(C)$ being a vector space complement. By construction, $C$ corestricts to an isomorphism of locally convex spaces onto its image. This shows the claimed statement.
\end{proof}

The proof of \Cref{Lem:ev} shows that the kernel of $\ev_0$ is co-Banach, whence $\ev_0$ is a submersion. If $\g$ is finite-dimensional, the proof can be shortened significantly. Moreover, in this case $\Gamma(\g)$ is a so-called Silva (or DFS-)space. However, we do not need the stronger properties of these spaces and instead recall from \cite[Theorem 5.10]{Glo04} (also confer the stronger results in \cite{DGS14}):
\begin{prop} \label{prop:LGPstruct_germs}
	Let $G$ be a (real) Banach Lie group with Lie algebra $\Lf(G)$ and Lie group exponential function $\exp_G \colon \Lf(G)\rightarrow G$. Then there is a uniquely determined real-analytic BCH Lie group structure on $\Gamma(G)$ modeled on $\Gamma(\Lf(G))$ such that the pushforward
	\begin{equation}
		(\exp_G)_* \colon \Gamma(\Lf(G)) \to \Gamma(G), [c] \mapsto [{\exp_G} \circ c]
	\end{equation}
	is a local diffeomorphism of real-analytic manifolds on some zero-neighborhood. The exponential function of $\Gamma(G)$ is this pushforward, $\exp_{\Gamma(G)} = (\exp_G)_*$.
\end{prop}

\begin{rem}
	There are versions of \Cref{prop:LGPstruct_germs} for germs of analytic mappings on compact subsets of metrizable topological vector spaces. However, in our setting we only need analytic curves, whence we chose to not present the results in full generality.
\end{rem}

\begin{lem} \label{lem:exp_inj_germs_through_zero}
	Let $G$ be a (real) Banach Lie group with Lie algebra $\g = \Lf(G)$ and exponential function $\exp_G$. Then the exponential function $\exp_{\Gamma(G)}$ of the germ group $\Gamma(G)$ is injective on the set of germs passing through zero
	\begin{equation}
		\{[c] \in \Gamma(\g) \mid c(0) = 0\} \, .
	\end{equation}
\end{lem}

\begin{proof}
	As Banach Lie groups are locally exponential, there is an open zero-neighborhood $V$ on which $\exp_G$ is injective. For germs passing through $0$, on a sufficiently small interval they take their image in $V$, whence injectivity on the germ level follows due to $\exp_{\Gamma(G)} ([c]) = [\exp_G \circ c]$.
\end{proof}

\section{Lie algebra contractions in terms of germ algebras} \label{sec:LAC-germ-algebras}

In this section, we formulate Lie algebra contractions and expansions in terms of germ algebras.

\begin{defn} \label{defn:Lie_alg_contr_germ}
	Let $\g$ be a Banach Lie algebra. Let $\phi \colon (-a,a) \to \Lin(\g,\g), \varepsilon \mapsto \phi_\varepsilon$ be an analytic curve defining an analytic Lie algebra contraction (in the sense of \Cref{defn:contr_analytic}) of $\g$.
	\begin{enumerate}[label=(\roman*)]
		\item We denote by $\Gamma(\g)_\phi \subseteq \Gamma(\g)$ the subset consisting of germs that arise by `pointwise application of $\phi$', i.e.
		\begin{equation}
			\Gamma(\g)_\phi := \{[c] \in \Gamma(\g) \mid \exists [\tilde{c}] \in \Gamma(\g) : c(\varepsilon) = \phi_\varepsilon(\tilde{c}(\varepsilon)) \} \, .
		\end{equation}
		This is a subalgebra of $\Gamma(\g)$ due to analyticity of $[ \, \cdot \, , \cdot \, ]_\varepsilon$: writing $[c_1], [c_2] \in \Gamma(\g)_\phi$ as $c_i(\varepsilon) = \phi_\varepsilon(\tilde{c}_i(\varepsilon))$ with $[\tilde{c}_i] \in \Gamma(\g)$, for their Lie bracket $[c] := [[c_1], [c_2]]_{\Gamma(\g)}$ we have
		\begin{align}
			c(\varepsilon) &= [c_1(\varepsilon), c_2(\varepsilon)]_\g \nonumber\\
			&= [\phi_\varepsilon(\tilde{c}_1(\varepsilon)), \phi_\varepsilon(\tilde{c}_2(\varepsilon))]_\g \nonumber\\
			&= \phi_\varepsilon([\tilde{c}_1(\varepsilon), \tilde{c}_2(\varepsilon)]_\varepsilon) \nonumber\\
			&= \phi_\varepsilon(\tilde{c}(\varepsilon))
		\end{align}
		in terms of $\tilde{c}(\varepsilon) := [\tilde{c}_1(\varepsilon), \tilde{c}_2(\varepsilon)]_\varepsilon$, which is analytic due to analyticity of $[ \, \cdot \, , \cdot \, ]_\varepsilon$.

		\item For any $k \in \N_0$, the set
			\begin{equation}
				\varepsilon^k \Gamma(\g)_\phi := \{\varepsilon^k [c] \mid [c] \in \Gamma(\g)_\phi\} = \{[\varepsilon \mapsto \varepsilon^k c(\varepsilon)] \mid [c] \in \Gamma(\g)_{\h} \} \subseteq \Gamma(\g)_\phi
			\end{equation}
			is a Lie ideal of $\Gamma(\g)_\phi$. The quotient
			\begin{equation}
				\g_\phi^{(k)} := \Gamma(\g)_\phi \Big/ \varepsilon^{k+1} \Gamma(\g)_\phi
			\end{equation}
			is the \emph{Lie algebra expansion of $\g$ with respect to $\phi$ to order $k$}.
	\end{enumerate}
\end{defn}

The expansion to order $0$ is isomorphic to the Lie algebra contraction with respect to $\phi$ in a natural way:

\begin{prop} \label{prop:Lie_alg_expansion_0_contr}
	Let $\g$ and $\phi$ be as above. Consider the continuous linear map
	\begin{subequations}
	\begin{equation}
		\Psi \colon \g_0 \to \Gamma(\g)_\phi
	\end{equation}
	defined by
	\begin{equation}
		\Psi(X) := [\varepsilon \mapsto \phi_\varepsilon(X)]
	\end{equation}
	\end{subequations}
	for any $X \in \g$ (as vector spaces, $\g$ and $\g_0$ agree). The induced map
	\begin{equation}
		\overline{\Psi} \colon \g_0 \to \g_\phi^{(0)} \, ,
	\end{equation}
	i.e.\ the composition of $\Psi$ with the quotient projection $\Gamma(\g)_\phi \to \g_\phi^{(0)}$, is a continuous Lie algebra isomorphism. (Note that we make no statement about continuity of its inverse; for a sufficient condition for continuity of the inverse, see \Cref{cons:expansion_0_contr_cont_inv}.)
\end{prop}

\begin{proof}
	By construction of $\Psi$, for any $X,Y \in \g$ the Lie bracket
	\begin{subequations} \label{eq:compute_germ_contr_1}
	\begin{align}
		[c] &:= [\Psi(X), \Psi(Y)]_{\Gamma(\g)_\phi} \nonumber\\
		& \, = \Big[[\varepsilon \mapsto \phi_\varepsilon(X)], [\varepsilon \mapsto \phi_\varepsilon(Y)]\Big]_{\Gamma(\g)} \\
		\shortintertext{is represented by}
		c(\varepsilon)
		&= [\phi_\varepsilon(X), \phi_\varepsilon(Y)]_\g \nonumber\\
		&= \phi_\varepsilon([X, Y]_\varepsilon).
	\end{align}
	\end{subequations}
	By analyticity, we have $[X, Y]_\varepsilon = [X,Y]_0 + \varepsilon \tilde{d}(\varepsilon)$ for some germ $[\tilde{d}] \in \Gamma(\g)$. Thus, we obtain
	\begin{subequations} \label{eq:compute_germ_contr_2}
	\begin{align}
		c(\varepsilon) &= \phi_\varepsilon([X,Y]_0) + \varepsilon \phi_\varepsilon(\tilde{d}(\varepsilon)) \nonumber\\
		&= \phi_\varepsilon([X,Y]_0) + \varepsilon d(\varepsilon),
	\end{align}
	where $d(\varepsilon) = \phi_\varepsilon (\tilde{d}(\varepsilon))$ defines a germ $[d] \in \Gamma(\g)_\phi$. Hence, we have
	\begin{equation}
		[\Psi(X), \Psi(Y)]_{\Gamma(\g)_\phi} = [c] = \Psi([X,Y]_0) + \varepsilon [d] \text{:}
	\end{equation}
	\end{subequations}
	the map $\Psi$ is a Lie algebra homomorphism $\g_0 \to \Gamma(\g)_\phi$ up to a term in $\varepsilon \Gamma(\g)_\phi$. Therefore, quotienting out $\varepsilon \Gamma(\g)_\phi$, the induced linear map $\overline{\Psi} \colon \g_0 \to \g_\phi^{(0)}$ is a Lie algebra homomorphism.

	So we are left to show invertibility of $\overline{\Psi}$. Surjectivity follows by a direct computation: for any element $\overline{[c]} = [c] + \varepsilon \Gamma(\g)_\phi \in \g_\phi^{(0)}$ with $[c] \in \Gamma(\g)_\phi$, there is a germ $[\tilde{c}] \in \Gamma(\g)$ such that $c(\varepsilon) = \phi_\varepsilon (\tilde{c}(\varepsilon))$. By analyticity we have $\tilde{c}(\varepsilon) = \tilde{c}(0) + \varepsilon d(\varepsilon)$ for some other germ $[d] \in \Gamma(\g)$. This implies that the value $\tilde{c}(0) \in \g$ is then a preimage of $\overline{[c]}$ under $\overline{\Psi}$: we have $c(\varepsilon) = \phi_\varepsilon(\tilde{c}(\varepsilon)) = \phi_\varepsilon (\tilde{c}(0)) + \varepsilon \phi_\varepsilon(d(\varepsilon))$, showing that $[c] = \Psi(\tilde{c}(0)) + [\varepsilon \mapsto \varepsilon \phi_\varepsilon(d(\varepsilon))]$, which implies $\overline{[c]} = \overline{\Psi}(\tilde{c}(0))$.

	For injectivity, let $X \in \g$ such that $\overline{\Psi}(X) = 0 \in \g_\phi^{(0)}$. By definition of $\overline{\Psi}$, this means $\Psi(X) \in \varepsilon \Gamma(\g)_\phi$. Hence there is a $[c] \in \Gamma(\g)_\phi$ such that $\Psi(X) = \varepsilon [c]$, i.e.\ $\phi_\varepsilon(X) = \varepsilon c(\varepsilon)$. By definition of $\Gamma(\g)_\phi$ there is a $[\tilde{c}] \in \Gamma(\g)$ such that $c(\varepsilon) = \phi_\varepsilon(\tilde{c}(\varepsilon))$. Thus we have $\phi_\varepsilon(X) = \varepsilon c(\varepsilon) = \varepsilon \phi_\varepsilon(\tilde{c}(\varepsilon)) = \phi_\varepsilon(\varepsilon \tilde{c}(\varepsilon))$. For $\varepsilon \ne 0$, the map $\phi_\varepsilon$ is injective, such that we obtain $X = \varepsilon \tilde{c}(\varepsilon)$. By analyticity, we may evaluate at $\varepsilon = 0$, yielding $X = 0$.

	We have thus shown that $\overline{\Psi}$ is bijective; hence it is a linear isomorphism, finishing the proof.
\end{proof}

\begin{cons} \label{cons:expansion_0_contr_cont_inv}
	In the general setting of \Cref{prop:Lie_alg_expansion_0_contr}, we cannot guarantee that $\overline{\Psi}$ is an isomorphism of locally convex spaces, i.e.\ that its inverse is continuous. Note that \emph{if} this holds, this makes $\overline{\Psi}$ an an isomorphism of locally convex Lie algebras, showing in particular that $\g_\phi^{(0)}$ is a Banach Lie algebra. In the following, we are going to provide a necessary condition for $\overline{\Psi}$ having a continuous inverse.

	The topology on $\Gamma(\g)$ is defined by considering it as real subspace of $(\Gamma(\g))_\C \cong \Gamma(\g_\C)$, which is topologized as the inductive limit (in the category of locally convex spaces) of the family of Banach spaces of bounded holomorphic functions $\mathrm{BHol} (B^\C_{1/n}(0), \g_\C)$ on balls in $\C$ of radius $1/n$ around $0$, $n \in \N$.

	Analogous to $\Gamma(\g)_\phi$, we can define $\mathrm{BHol} (B^\C_{1/n}(0), \g_\C)_\phi$ as the space of those bounded holomorphic $\g_\C$-valued functions on $B^\C_{1/n}(0)$ which arise by pointwise application of $\phi_\varepsilon$ to an element of $\mathrm{BHol} (B^\C_{1/n}(0), \g_\C)$. Analogous to $\Psi$, we can further define maps $\Psi^\C_{1/n} \colon \g_\C \to \mathrm{BHol} (B^\C_{1/n}(0), \g_\C)_\phi$. The arguments from the proof of \Cref{prop:Lie_alg_expansion_0_contr} then show that the induced maps
	\begin{equation} \label{eq:expansion_0_contr_cont_inv}
		\overline{\Psi}^\C_{1/n} \colon \g_\C \to \mathrm{BHol} (B^\C_{1/n}(0), \g_\C)_\phi \Big/ \varepsilon \mathrm{BHol} (B^\C_{1/n}(0), \g_\C)_\phi
	\end{equation}
	are continuous linear isomorphisms. If we now \emph{assume that the spaces $\mathrm{BHol} (B^\C_{1/n}(0), \g_\C)_\phi \subseteq \mathrm{BHol} (B^\C_{1/n}(0), \g_\C)$ are closed}, then all involved spaces are (complex) Banach spaces, such that by the open mapping theorem the inverse of $\overline{\Psi}^\C_{1/n}$ is continuous as well. Now since both inductive limits and quotients are colimits in the categorical sense, and colimits commute, the quotient $(\g_\phi^{(0)})_\C$ of the limits is the limit of the quotients. But the natural map (induced by restriction) between two of the quotients $\mathrm{BHol} (B^\C_{1/n_i}(0), \g_\C)_\phi \Big / \varepsilon \mathrm{BHol} (B^\C_{1/n_i}(0), \g_\C)_\phi$ for $n_1 < n_2$ is an isomorphism as well. Hence, taking the inductive limit of \eqref{eq:expansion_0_contr_cont_inv}, the induced map
	\begin{equation}
		\overline{\Psi}^\C \colon \g_\C \to (\g_\phi^{(0)})_\C
	\end{equation}
	is an isomorphism of complex locally convex spaces. By construction, it is $\R$-linear, and its restriction to the real subspaces is $\overline{\Psi}$, showing that $\overline{\Psi}$ is an isomorphism of (real) locally convex spaces.

	Of course, in general it is not obvious to check if a given family $\varepsilon \mapsto \phi_\varepsilon$ satisfies the condition that the spaces $\mathrm{BHol} (B^\C_{1/n}(0), \g_\C)_\phi \subseteq \mathrm{BHol} (B^\C_{1/n}(0), \g_\C)$ be closed (which would in particular also imply that $\Gamma(\g)_\phi \subseteq \Gamma(\g)$ is closed). But for example, this condition is satisfied in the \.{I}nönü--Wigner special case as discussed in \Cref{cons:Inonu--Wigner} below.
\end{cons}

\begin{rem} \label{rem:higher-expansions-equivalence}
	\begin{enumerate}[label=(\roman*)]
		\item By an argument analogous to the proof of \Cref{prop:Lie_alg_expansion_0_contr}, for each $k \in \N$ the continuous linear map
			\begin{equation}
				\Psi^{(k)} \colon \g^{k+1} = \g \times \dots \times \g \ni (X_0, \dots, X_k) \mapsto \left [ \varepsilon \mapsto \phi_\varepsilon \left ( \sum_{\ell=0}^k \varepsilon^\ell X_\ell \right ) \right ] \in \Gamma(\g)_\phi
			\end{equation}
			induces a continuous linear isomorphism $\overline{\Psi^{(k)}} \colon \g^{k+1} \xrightarrow{\cong} \g_\phi^{(k)}$.

			As in \Cref{cons:expansion_0_contr_cont_inv}, assuming that the spaces $\mathrm{BHol} (B^\C_{1/n}(0), \g_\C)_\phi \subseteq \mathrm{BHol} (B^\C_{1/n}(0), \g_\C)$ are closed guarantees that $\overline{\Psi^{(k)}}$ has a continuous inverse.
		\item The higher-order expansions $\g_\phi^{(k)}$ encode `how the contraction is approached in the limit $\varepsilon \to 0$': instead of quotienting out the linear terms in $\varepsilon$ of $[\cdot, \cdot]_\varepsilon$ as in the computation of the order-$0$ expansion in \eqref{eq:compute_germ_contr_1}, \eqref{eq:compute_germ_contr_2}, at order $k > 0$ we keep all terms up to order $\varepsilon^k$.

			Therefore, in some sense, the space of germs $\Gamma(\g)_\phi$ without any quotienting encodes how the contraction is approached `to arbitrarily high order'.
	\end{enumerate}
\end{rem}

\begin{cons} \label{cons:Inonu--Wigner}
	In the \.{I}nönü--Wigner special case, i.e.\ $\g = \h \oplus \n$ (as a Banach space) and $\phi_\varepsilon(X_\h + X_\n) = X_\h + \varepsilon X_\n$, we obtain
	\begin{equation}
		\Gamma(\g)_\phi = \{ [c] \in \Gamma(\g) \mid c(0) \in \h\}.
	\end{equation}
	In particular, this shows that the contraction is independent of the choice of $\n$.

	Hence, when applying our previous construction of Lie algebra expansions to the \.{I}nönü--Wigner case, we can consider the more general situation where $\h$ is not necessarily complemented, as we will do in the following.
\end{cons}

\begin{defn}
	Let $\h$ be a Lie subalgebra of the Banach Lie algebra $\g$ (mostly, we will be interested in the case where $\h$ is closed). We write
	\begin{equation}
		\Gamma(\g)_\h := \{[c] \in \Gamma(\g) \mid c(0) \in \h\}
	\end{equation}
	for the \emph{Lie subalgebra of all germs passing through} $\h$.

	For $k \in \N_0$, we have a Lie ideal $\varepsilon^k \Gamma(\g)_\h \trianglelefteq \Gamma(\g)_\h$. The quotient
	\begin{equation}
		\g_\h^{(k)} := \Gamma(\g)_\h \Big/ \varepsilon^{k+1} \Gamma(\g)_\h
	\end{equation}
	is the \emph{Lie algebra expansion of $\g$ with respect to $\h$ to order $k$.}
\end{defn}

\begin{lem}
	If $\h$ is a closed (or complemented) Lie subalgebra of the Banach Lie algebra $\g$, then $\Gamma(\g)_\h$ is a closed (or complemented and co-Banach) subalgebra of $\Gamma(\g)$.
\end{lem}

\begin{proof}
	As the Lie bracket is computed pointwise, it is clear that $\Gamma(\g)_{\h}$ is a Lie subalgebra and we only need to establish the topological properties.

	The point evaluation map $\ev_0$ is continuous linear by \Cref{Lem:ev}. Hence if $\h$ is closed, the preimage $\Gamma(\g)_{\h}=\ev_0^{-1}(\h)$ is a closed Lie subalgebra. If $\h$ is in addition complemented, then there is a continuous projection $\pi_\h \colon \g \rightarrow \g$, $\pi_\h \circ \pi_\h = \pi_\h$, such that $\pi_\h(\g) = \h$. Now consider
	\begin{equation}
		\pi_{\Gamma(\g)_\h} \colon \Gamma(\g) \to \Gamma(\g), \quad [c] \mapsto [c] - [\varepsilon \mapsto c(0) - \pi_\h(c(0))] \, .
	\end{equation}
	Applying \Cref{Lem:ev}, we see that $\pi_{\Gamma(\g)_\h} = \id_{\Gamma(\g)} - C \circ (\id_\g -\pi_\h) \circ \ev_0$ is continuous linear. Note that for every $[c] \in \Gamma(\g)_\h$ we have $\pi_{\Gamma(\g)_\h}([c]) = [c]$. Now as $\pi_{\h}$ is a continuous projection, it is easy to see that $\pi_{\Gamma(\g)_\h}$ is a continuous projection. We conclude that $\Gamma(\g)_\h$ is a complemented Lie subalgebra. The complement is given by $C(\ker\pi_\h)$, which is a closed subspace of the Banach space $C(\g) \subseteq \Gamma(\g)$, and thus $\Gamma(\g)_\h$ is co-Banach.
\end{proof}

\begin{prop}
	If $\h$ is a closed Lie subalgebra of the Banach Lie algebra $\g$, then the ideals $\varepsilon^k \Gamma(\g)_\h \trianglelefteq \Gamma(\g)_\h$ are closed.
\end{prop}

\begin{proof}
	We can write the ideals as $\varepsilon^k \Gamma(\g)_\h = \{ [c] \mid c(0) = c'(0) = \dots = c^{(k-1)}(0) = 0, c^{(k)}(0) \in \h \}$, and evaluation of derivatives of any order is continuous by \Cref{Lem:ev}.
\end{proof}

\section{Lie group contractions in germ groups} \label{sec:LGC-germ-groups}

In this section we consider the global version of the quotienting process giving rise to Lie algebra expansions, considered in the previous section. For technical reasons, we will only cover the \.{I}nönü--Wigner special case, leaving the general case for future work.

The setting in this section will be as follows:
\begin{setup}
	Throughout this section $G$ will be a Banach Lie group with Lie algebra $\g = \Lf(G)$. Further $H \subseteq G$ will be a Lie subgroup with Lie algebra $\h = \Lf(H) \subseteq \g$.
\end{setup}

Our goal is to show that the construction of the Lie algebras $\Gamma(\g)_\h \subseteq \Gamma(\g)$, $\varepsilon^k \Gamma(\g)_\h \trianglelefteq \Gamma(\g)_\h$, and $\g_\h^{(k)} = \Gamma(\g)_\h \Big / \varepsilon^{k+1} \Gamma(\g)_\phi$ can be `lifted' to the Lie group level.

\begin{lem} \label{lem:germs_subgroup}
	The group
	\begin{equation}
		\Gamma(G)_H := \{[f] \in \Gamma(G) \mid f(0) \in H\}
	\end{equation}
	is an analytic Lie subgroup of $\Gamma(G)$, with Lie algebra $\Lf(\Gamma(G)_H) = \Gamma(\g)_\h$.
\end{lem}

\begin{proof}
	First, we note that $\Gamma(G)_H \subseteq \Gamma(G)$ is closed, since point evaluation is continuous and $H \subseteq G$ is closed.

	Quite directly, one obtains
	\begin{equation} \label{eq:pf_germs_subgroup_Lie_alg}
		\Lf^e(\Gamma(G)_H) = \Gamma(\g)_\h \, \text{:}
	\end{equation}
	For the inclusion `$\supseteq$', let $[c] \in \Gamma(\g)_\h$. Applying the exponential function, we have $\exp_{\Gamma(G)}([c]) = [{\exp_G} \circ c]$, which evaluated at $0$ yields $\bigl ( \exp_{\Gamma(G)}([c]) \bigr ) (0) = \exp_G(c(0))$. Since $c(0) \in \h$, we have $\exp_G(c(0)) \in H$. This shows the desired inclusion. Conversely, for the inclusion `$\subseteq$', consider a $[c] \in \Lf^e(\Gamma(G)_H)$. In particular, this implies $\exp_{\Gamma(G)}([c]) \in \Gamma(G)_H$, i.e.\ $[{\exp_G} \circ c] \in \Gamma(G)_H$. This means $\exp_G(c(0)) \in H$, which implies $c(0) \in \h$, showing $[c] \in \Gamma(\g)_\h$ as desired.

	We now want to apply \Cref{thm:Lie_subgroup} to show that $\Gamma(G)_H$ is an analytic Lie subgroup of $\Gamma(G)$. Hence, we need to show that there is an open zero-neighborhood $V \subseteq \Gamma(\g)$ such that ${\exp_{\Gamma(G)}}|_V$ is an injective map onto an open set in $\Gamma(G)$ and
	\begin{equation} \label{eq:pf_germs_subgroup_exp_coords}
		\exp_{\Gamma(G)}(V \cap \Gamma(\g)_\h) = \exp_{\Gamma(G)}(V) \cap \Gamma(G)_H
	\end{equation}
	(here we have already used \eqref{eq:pf_germs_subgroup_Lie_alg}). In fact, we are going to show a stronger statement: we will show that \eqref{eq:pf_germs_subgroup_exp_coords} holds for \emph{any} open zero-neighborhood $V \subseteq \Gamma(\g)$ such that ${\exp_{\Gamma(G)}}|_V$ is an injective map onto an open set in $\Gamma(G)$ (which exists due to $\Gamma(G)$ being locally exponential).

	So let $V$ be such an open set. Note that the inclusion `$\subseteq$' in \eqref{eq:pf_germs_subgroup_exp_coords} holds by \eqref{eq:pf_germs_subgroup_Lie_alg}, such that we only need to show the inclusion `$\supseteq$'. For this, consider an $[f] \in \exp_{\Gamma(G)}(V) \cap \Gamma(G)_H$. Since it is in $\exp_{\Gamma(G)}(V)$, there is a unique $[c] \in V$ with $[f] = \exp_{\Gamma(G)}([c])$, i.e.\ $f = {\exp_G} \circ c$. We also know $[f] \in \Gamma(G)_H$, which means $f(0) \in H$; hence we have $\exp_G(c(0)) \in H$. This implies $c(0) \in \h$. Hence we have $[c] \in \Gamma(\g)_\h$, finishing the proof.
\end{proof}

\begin{lem} \label{lem:expansion_subgroup}
	For $k \in \N$, consider the subset
	\begin{equation}
		\Gamma(G)_H^{(k)} := \{[{\exp_G} \circ c] \mid [c] \in \varepsilon^k \Gamma(\g)_\h\} = \exp_{\Gamma(G)}(\varepsilon^k \Gamma(\g)_\h)
	\end{equation}
	of $\Gamma(G)$. This is an analytic Lie subgroup of $\Gamma(G)$, and hence (by \Cref{lem:germs_subgroup}) of $\Gamma(G)_H$. In $\Gamma(G)_H$, it is normal.
\end{lem}

\begin{proof}
	First, we show that $\Gamma(G)_H^{(k)} \subseteq \Gamma(G)$ is closed. Note that $\Gamma(G)_H^{(k)}$ can be written as the set of all analytic germs in $\Gamma(G)$ which evaluate to $\mathbf{1}_G$ at $0$ and whose first $k-1$ derivatives at $0$, in exponential coordinates, are $0$ (i.e.\ whose $(k-1)$-jet\footnote{For a discussion of jets in the context of Banach manifolds, see \cite[Section 9.1]{MROD92}.} at $0$ is equal to that of the constant curve $\varepsilon \mapsto \mathbf{1}_G$), and whose $k$-th derivative is in $\h$, i.e.\ we have
	\begin{equation} \label{eq:pf_expansion_subgroup_closed}
		\Gamma(G)_H^{(k)} = \ev_0^{-1}(\mathbf{1}_G) \cap \left ( \bigcap_{0 < \ell \le k-1}(\ev_0 \circ D^\ell)^{-1}(0) \right ) \cap (\ev_0 \circ D^k)^{-1}(\mathfrak{h}) \, .
	\end{equation}
	Since $\Gamma(G)$ is topologized analogously to $\Gamma(\g)$---namely as a closed real subspace of a space of germs of holomorphic $G$-valued mappings, which is an inductive limit of spaces of bounded holomorphic maps defined on discs in $\C$ around $0$ \cite[Appendix~A]{DS15}---the point evaluation map $\ev_0$ and the derivative operators $D^\ell$ may be shown to be continuous in exact analogy to the $\Gamma(\g)$ case in \Cref{Lem:ev}. Hence, \eqref{eq:pf_expansion_subgroup_closed} expresses $\Gamma(G)_H^{(k)}$ as an intersection of closed sets, thus showing that it is closed.

	Next, we show that $\Gamma(G)_H^{(k)} \subseteq \Gamma(G)$ is a subgroup. This is due to Banach Lie groups being BCH, and us being able to shrink the definition interval of the germs: there is an open zero-neighborhood $A \subseteq \g$ such that on $A \times A$, the BCH series converges to an analytic \emph{local multiplication} map $* \colon A \times A \to \g$, and this satisfies
	\begin{equation}
		\exp_G(X) \exp_G(Y) = \exp_G(X * Y) \; \text{for all} \; X, Y \in A \, .
	\end{equation}
	Now given $[c], [d] \in \varepsilon^k \Gamma(\g)_\h$, we have $c(0) = 0 = d(0)$, such that for $\varepsilon$ small enough we have $c(\varepsilon), d(\varepsilon) \in A$. This implies that for these $\varepsilon$, we have $\exp_G(c(\varepsilon)) \exp_G(d(\varepsilon)) = \exp_G(c(\varepsilon) * d(\varepsilon))$. This shows that
	\begin{equation} \label{eq:pf_expansion_subgroup}
		\exp_{\Gamma(G)}([c]) \exp_{\Gamma(G)}([d]) = \exp_{\Gamma(G)}([c * d]) \, ,
	\end{equation}
	where we have defined $(c * d)(\varepsilon) := c(\varepsilon) * d(\varepsilon)$. Due to the terms of the BCH series being homogeneous polynomials, when expressing $c(\varepsilon)$ and $d(\varepsilon)$ as power series in the computation of $c * d$, the powers of $\varepsilon$ `pile up', such that the resulting germ $[c * d]$ lies again in $\varepsilon^k \Gamma(\g)_\h$. Thus, \eqref{eq:pf_expansion_subgroup} shows that $\Gamma(G)_H^{(k)} = \exp_{\Gamma(G)}(\varepsilon^k \Gamma(\g)_\h)$ is indeed a subgroup.

	Now, we are going to show
	\begin{equation} \label{eq:pf_expansion_subgroup_Lie_alg}
		\Lf^e(\Gamma(G)_H^{(k)}) = \varepsilon^k \Gamma(\g)_\h \, .
	\end{equation}
	The inclusion `$\supseteq$' holds by definition of $\Gamma(G)_H^{(k)}$. For the inclusion `$\subseteq$', consider any $[c] \in \Lf^e(\Gamma(G)_H^{(k)})$. This means that for all $\lambda \in \R$ we have $\exp_{\Gamma(G)}(\lambda [c]) \in \Gamma(G)_H^{(k)}$. In particular, this implies that $\exp_{\Gamma(G)}(\lambda [c])$ evaluates to $\mathbf{1}_G$ at $0$ for all $\lambda \in \R$, i.e.\ $\mathbf{1}_G = \exp_G(\lambda c(0))$ for all $\lambda \in \R$. This shows that $c(0) = 0$. On the other hand, we have $\exp_{\Gamma(G)}([c]) \in \Gamma(G)_H^{(k)} = \exp_{\Gamma(G)}(\varepsilon^k \Gamma(\g)_\h)$. Therefore, there is a germ $[d] \in \varepsilon^k \Gamma(\g)_\h$ with
	\begin{equation}
		\exp_{\Gamma(G)}([c]) = \exp_{\Gamma(G)}([d]).
	\end{equation}
	From $[d] \in \varepsilon^k \Gamma(\g)_\h$, we get in particular that $d(0) = 0$. Hence, injectivity of $\exp_{\Gamma(G)}$ on germs passing through zero (\Cref{lem:exp_inj_germs_through_zero}) implies $[c] = [d]$. This shows the desired inclusion.

	As in the proof of \Cref{lem:germs_subgroup}, we now want to apply \Cref{thm:Lie_subgroup} to show that $\Gamma(G)_H^{(k)}$ is an analytic Lie subgroup of $\Gamma(G)$. Hence, we need to show that there is an open zero-neighborhood $V \subseteq \Gamma(\g)$ such that ${\exp_{\Gamma(G)}}|_V$ is an injective map onto an open set in $\Gamma(G)$ and
	\begin{equation} \label{eq:pf_expansion_subgroup_exp_coords}
		\exp_{\Gamma(G)}(V \cap \varepsilon^k \Gamma(\g)_\h) = \exp_{\Gamma(G)}(V) \cap \Gamma(G)_H^{(k)} \, .
	\end{equation}

	Since $G$ is a Banach Lie group, it is in particular locally exponential. Therefore, there is an open zero-neighborhood $\tilde{V} \subseteq \g$ such that $\tilde{U} := \exp_G(\tilde{V})$ is open and ${\exp_G}|_{\tilde{V}}^{\tilde{U}}$ is a diffeomorphism. Consider the sets $V := (\ev_0^\g)^{-1}(\tilde{V}) \subseteq \Gamma(\g)$, $U := (\ev_0^G)^{-1}(\tilde{U}) \subseteq \Gamma(G)$. Since point evaluation of germs is continuous, these are open. By definition, we have $[\varepsilon \mapsto 0] \in V$ and $U = \exp_{\Gamma(G)}(V)$. Further, ${\exp_{\Gamma(G)}}|_V$ is injective: consider two germs $[c], [d] \in V$ with $\exp_{\Gamma(G)}([c]) = \exp_{\Gamma(G)}([d])$, i.e.\ $\exp_G(c(\varepsilon)) = \exp_G(d(\varepsilon))$ for $\varepsilon$ in the intersection of the domains. By definition of $V$, we have $c(0), d(0) \in \tilde{V}$. Continuity of $c, d$ implies that for $\varepsilon$ sufficiently small, we have $c(\varepsilon), d(\varepsilon) \in \tilde{V}$ as well. But $\exp_G$ is injective on $\tilde{V}$, such that we obtain $c(\varepsilon) = d(\varepsilon)$ for $\varepsilon$ small enough, i.e.\ $[c] = [d]$.

	Now we will prove that \eqref{eq:pf_expansion_subgroup_exp_coords} holds for this $V$. The inclusion `$\subseteq$' holds by \eqref{eq:pf_expansion_subgroup_Lie_alg}, so we only need to show the inclusion `$\supseteq$'. Consider an $[f] \in \exp_{\Gamma(G)}(V) \cap \Gamma(G)_H^{(k)}$. This means that there are $[c] \in V$ and $[d] \in \varepsilon^k \Gamma(\g)_\h$ with $\exp_{\Gamma(G)}([c]) = [f] = \exp_{\Gamma(G)}([d])$. Since $f(0) = \mathbf{1}_G$, we have $\exp_G(c(0)) = f(0) = \mathbf{1}_G = \exp_G(0)$. By definition of $V$, we have $c(0) \in \tilde{V}$; since $0 \in \tilde{V}$ and $\exp_G$ is injective on $\tilde{V}$, we obtain $c(0) = 0$. Hence, due to injectivity of $\exp_{\Gamma(G)}$ on germs passing through zero (\Cref{lem:exp_inj_germs_through_zero}) we have $[c] = [d]$, proving the inclusion.

	Finally, we are going to show that $\Gamma(G)_H^{(k)}$ is a normal subgroup of $\Gamma(G)_H$. First, we note that it being normal in the \emph{identity component} of $\Gamma(G)_H$ follows from $\Lf(\Gamma(G)_H^{(k)}) = \varepsilon^k \Gamma(\g)_\h$ being a closed Lie ideal in $\Lf(\Gamma(G))_H = \Gamma(\g)_\h$: since $\Gamma(G)_H$ is locally exponential, the finite-dimensional argument \cite[Proposition 1.11.7]{DK00} carries over verbatim.

	In order to extend this to the whole of $\Gamma(G)_H$, we proceed as follows. Any element $[f] \in \Gamma(G)_H$ can be written as $[f] = [\varepsilon \mapsto h \cdot g(\varepsilon)]$ where $h \in H$ and $g$ is a $G$-valued analytic curve satisfying $g(0) = \mathbf{1}_G$ (indeed, we may set $h := f(0)$ and $g(\varepsilon) := h^{-1} f(\varepsilon)$). In particular, the germ $[g]$ lies in the identity component of $\Gamma(G)_H$, such that the previous argument shows that $\Gamma(G)_H^{(k)}$ is invariant under conjugation by $[g]$. Hence, we are left to show invariance of $\Gamma(G)_H^{(k)}$ under conjugation by constant $H$-valued germs. However, this is immediate: given any $h \in H$ and $[{\exp_G} \circ c] \in \Gamma(G)_H^k$, we have
	\begin{subequations}
	\begin{align}
		h \cdot \exp_G(c(\varepsilon)) \cdot h^{-1} &= \exp_G(\mathrm{Ad}_h(c(\varepsilon))), \\
		\shortintertext{i.e.}
		[h] \cdot [{\exp_G} \circ c] \cdot [h]^{-1} &= [{\exp_G} \circ (\mathrm{Ad}_h \circ c)] \, ;
	\end{align}
	\end{subequations}
	and due to $\mathrm{Ad}_h$ being linear and leaving $\h \subseteq \g$ invariant, $[c] \in \varepsilon^k \Gamma(\g)_\h$ implies that $[\mathrm{Ad}_h \circ c] \in \varepsilon^k \Gamma(\g)_\h$.
\end{proof}

Combined, we arrive at our first main result, the desired lifting of Lie algebra expansions to the group level:
\begin{thm} \label{thm:main_result}
	Let $G$ be a Banach Lie group with Lie algebra $\g = \Lf(G)$, and $H \subseteq G$ a Lie subgroup with Lie algebra $\h = \Lf(H)$.
	\begin{enumerate}[label=(\roman*)]
		\item The group $\Gamma(G)_H$ is an analytic Lie subgroup of $\Gamma(G)$, with Lie algebra $\Lf(\Gamma(G)_H) = \Gamma(\g)_\h$.

		\item For any $k \in \N_0$, we have that
			\begin{equation}
				\Gamma(G)_H^{(k+1)} := \{[{\exp_G} \circ c] \mid [c] \in \varepsilon^{k+1} \Gamma(\g)_\h\} = \exp_{\Gamma(G)}(\varepsilon^{k+1} \Gamma(\g)_\h)
			\end{equation}
			is a normal analytic Lie subgroup of $\Gamma(G)_H$, with Lie algebra $\Lf(\Gamma(G)_H^{(k+1)}) = \varepsilon^{k+1} \Gamma(\g)_\h$.

			Hence, the quotient
			\begin{equation}
				G^{(k)}_H := \Gamma(G)_H \Big/ \Gamma(G)_H^{(k+1)}
			\end{equation}
			is a BCH Lie group with Lie algebra $\Lf(G^{(k)}_H) = \g^{(k)}_\h$.
	\end{enumerate}
\end{thm}

\begin{proof}
	This is a combination of \Cref{lem:germs_subgroup,lem:expansion_subgroup}, and the quotient theorem for BCH Lie groups (\Cref{thm:quotient}).
\end{proof}

In the following, we are going to determine the topological structure of the `Lie group expansions' $G^{(k)}_H$ from \Cref{thm:main_result}. First, we note the following:
\begin{lem} \label{lem:germ_H_decomp}
	Any $[f] \in \Gamma(G)_H$ can be uniquely written in the form
	\begin{equation}
		[f] = [{\exp_G} \circ c] \cdot [\varepsilon \mapsto h]
	\end{equation}
	with $h \in H$ and $[c] \in \Gamma(\g)$ with $c(0) = 0$.
\end{lem}

\begin{proof}
	Defining $h = f(0)$, we have $f(\varepsilon) = g(\varepsilon) \cdot h$, where $[g] \in \Gamma(G)$ satisfies $g(0) = \mathbf{1}_G$. For $\varepsilon$ sufficiently small, $g(\varepsilon)$ lies in an open $\mathbf{1}_G$-neighborhood in $G$ onto which $\exp_G$ is a diffeomorphism, such that $g$ can be uniquely written as $[g] = [{\exp_G} \circ c]$ as desired.
\end{proof}

\begin{cons}
	Let $k \in \N_0$, and consider $[f], [g] \in \Gamma(G)_H$. These represent the same element in $G^{(k)}_H = \Gamma(G)_H \Big / \Gamma(G)_H^{(k+1)}$ if and only if $[g^{-1} \cdot f] \in \Gamma(G)_H^{(k+1)}$. In this case, we have in particular $\mathbf{1}_G = (g^{-1} \cdot f)(0)$, i.e.\ $f(0) = g(0) =: h \in H$.

	According to \Cref{lem:germ_H_decomp}, we now write the two germs as $[f] = [{\exp_G} \circ c] \cdot [\varepsilon \mapsto h]$, $[g] = [{\exp_G} \circ d] \cdot [\varepsilon \mapsto h]$ with $[c], [d] \in \Gamma(\g)$, $c(0) = d(0) = 0$. For $\varepsilon$ sufficiently small, we then have
	\begin{align}
		(g^{-1} \cdot f)(\varepsilon) &= h^{-1} \cdot \exp_G(-d(\varepsilon)) \cdot \exp_G(c(\varepsilon)) \cdot h \nonumber\\
		&= \exp_G\Big(\mathrm{Ad}_{h^{-1}}\big((-d(\varepsilon)) * c(\varepsilon)\big)\Big) \, ,
	\end{align}
	where $*$ denotes local multiplication, i.e.\ the analytic map to which the BCH series converges on an open zero-neighborhood in $\g$. This shows that $[f], [g]$ represent the same element in $G^{(k)}_H$ if and only if $[\mathrm{Ad}_{h^{-1}} \circ ((-d) * c)] \in \varepsilon^{k+1} \Gamma(\g)_\h$, i.e.\ if and only if $[(-d) * c] \in \varepsilon^{k+1} \Gamma(\g)_\h$.

	We now express $c,d$ via their power series expansions as $c(\varepsilon) = \sum_{\ell=1}^\infty \varepsilon^\ell c_{(\ell)}$, $d(\varepsilon) = \sum_{\ell=1}^\infty \varepsilon^\ell d_{(\ell)}$. Inserting these expansions into the BCH series and considering the coefficients of powers $\varepsilon^\ell$ of increasing order $1 \le \ell \le k+1$, we see that $[(-d) * c] \in \varepsilon^{k+1} \Gamma(\g)_\h$ if and only if
	\begin{equation}
		c_{(\ell)} = d_{(\ell)} \; \text{for} \; 1 \le \ell \le k \quad \text{and} \quad c_{(k+1)} - d_{(k+1)} \in \h \, .
	\end{equation}

	We thus have shown that there is a well-defined bijection
	\begin{subequations} \label{eq:map_topology_group_expansion}
	\begin{align}
		H \times \g^k \times (\g/\h) & \xrightarrow{\cong} G^{(k)}_H \, , \\
		(h, c_{(1)}, \dots, c_{(k)}, c_{(k+1)} + \h) &\mapsto \left[ \varepsilon \mapsto \exp_G \left ( \sum_{\ell=1}^{k+1} \varepsilon^\ell c_{(\ell)} \right ) \cdot h \right] \cdot \Gamma(G)_H^{(k+1)} \, .
	\end{align}
	\end{subequations}

	The map \eqref{eq:map_topology_group_expansion} is continuous, since it is induced by the continuous map
	\begin{equation}
		H \times \g^{k+1} \ni (h, c_{(1)}, \dots, c_{(k+1)}) \mapsto \left [ \varepsilon \mapsto \exp_G \left ( \sum_{\ell=1}^{k+1} \varepsilon^\ell c_{(\ell)} \right ) \cdot h \right ] \in \Gamma(G)_H \, .
	\end{equation}
	The inverse of \eqref{eq:map_topology_group_expansion} is induced by
	\begin{equation}
		\Gamma(G)_H \ni \left [ \varepsilon \mapsto \exp_G \left(\sum_{\ell=1}^\infty \varepsilon^\ell c_{(\ell)} \right ) \cdot h \right ] \mapsto (h, c_{(1)}, \dots, c_{(k+1)}) \in H \times \g^{k+1} \, .
	\end{equation}
	This may be expressed via the point evaluation map and derivative operators on $\Gamma(G)$, which are continuous (as discussed in the proof of \Cref{lem:expansion_subgroup}), and hence is continuous as well. Therefore, the map \eqref{eq:map_topology_group_expansion} is actually a homeomorphism.
\end{cons}

\begin{cons}
	Now, we are going to compute what the group multiplication of $G^{(k)}_H$ looks like when `pulled back' to $H \times \g^k \times (\g/\h)$ via the map \eqref{eq:map_topology_group_expansion}.

	First, we determine what the group multiplication of $\Gamma(G)_H$ looks like in terms of the decomposition of elements given by \Cref{lem:germ_H_decomp}. Consider two elements $[f], [g] \in \Gamma(G)_H$ written as $[f] = [{\exp_G} \circ c] \cdot [\varepsilon \mapsto h]$, $[g] = [{\exp_G} \circ d] \cdot [\varepsilon \mapsto \tilde{h}]$ with $[c], [d] \in \Gamma(\g)$, $c(0) = 0 = d(0)$, and $h, \tilde{h} \in H$. Their product is
	\begin{align}
		[f] \cdot [g] &= [\varepsilon \mapsto \exp_G(c(\varepsilon)) \cdot h \cdot \exp_G(d(\varepsilon)) \cdot \tilde{h}] \nonumber\\
		&= [\varepsilon \mapsto \exp_G(c(\varepsilon)) \cdot h \cdot \exp_G(d(\varepsilon)) \cdot h^{-1} \cdot h \cdot \tilde{h}] \nonumber\\
		&= [\varepsilon \mapsto \exp_G(c(\varepsilon)) \cdot \exp_G(\mathrm{Ad}_h(d(\varepsilon))) \cdot h \cdot \tilde{h}] \nonumber\\
		&= [{\exp_G} \circ (c * (\mathrm{Ad}_h \circ d))] \cdot [\varepsilon \mapsto h \tilde{h}]
	\end{align}

	We can thus determine the group structure induced on the topological space $H \times \g^k \times (\g/\h)$ by pulling back that of $G^{(k)}_H$ via \eqref{eq:map_topology_group_expansion}. It is a semidirect product
	\begin{equation} \label{eq:pullback_group_expansion}
		H \ltimes_\mathrm{Ad} (\g^k \times (\g/\h), \circledast) \, ,
	\end{equation}
	where the group multiplication $\circledast$ on $\g^k \times (\g/\h)$ arises from the truncation of the local multiplication map $*$ applied to polynomials in $\varepsilon$.

	Explicitly, this means that $\circledast$ is determined as follows. Given
	\begin{equation}
		(c_{(1)}, \dots, c_{(k)}, c_{(k+1)} + \h), (d_{(1)}, \dots, d_{(k)}, d_{(k+1)} + \h) \in \g^k \times (\g/\h) \, ,
	\end{equation}
	we consider the associated curves $c(\varepsilon) = \sum_{\ell=1}^{k+1} \varepsilon^\ell c_{(\ell)}, d(\varepsilon) = \sum_{\ell=1}^{k+1} \varepsilon^\ell d_{(\ell)}$. Of these, we compute the product $c(\varepsilon) * d(\varepsilon)$ with respect to the local multiplication, which has an expansion as a power series $c(\varepsilon) * d(\varepsilon) =: \sum_{\ell=1}^\infty \varepsilon^\ell (c*d)_{(\ell)}$. The first $k+1$ coefficients in this expansion then give
	\begin{equation} \label{eq:nustar}
		(c_{(1)}, \dots, c_{(k)}, c_{(k+1)} + \h) \circledast (d_{(1)}, \dots, d_{(k)}, d_{(k+1)} + \h) = ((c*d)_{(1)}, \dots, (c*d)_{(k)}, (c*d)_{(k+1)} + \h) \, .
	\end{equation}
	By construction, we know that this yields a well-defined element of $\g^k \times (\g/\h)$.

	Note that for the computation of the order-$k$ $\circledast$ operation, we only need to know the resulting power series expansion of $c(\varepsilon) * d(\varepsilon)$ to order $\varepsilon^{k+1}$, such that only finitely many terms of the BCH series are needed. Since the terms in the BCH series are (homogeneous) polynomials, this shows that $\circledast$ is in particular analytic when understood as a map between Banach spaces
	\begin{equation}
		\circledast \colon (\g^k \times (\g/\h)) \times (\g^k \times (\g/\h)) \to \g^k \times (\g/\h) \, .
	\end{equation}
	Furthermore, inverse elements with respect to $\circledast$ are simply given by taking the negative, which is also analytic.

	Thus, we have shown that the induced group structure \eqref{eq:pullback_group_expansion} on $H \times \g^k \times (\g/\h)$, understood as a Banach manifold, makes it into a Banach Lie group (using that the adjoint representation is analytic). Hence, the homeomorphism \eqref{eq:map_topology_group_expansion} is a continuous isomorphism of Banach Lie groups. As Banach Lie groups are BCH, the automatic smoothness theorem (\cite[Proposition 2.4]{Glo02}, \cite[Theorem IV.1.18]{Neeb06}) applies and shows that the map is an analytic diffeomorphism. Hence, we have proved the following, which is our second main result:
\end{cons}

\begin{thm} \label{thm:k-th-expansion}
	Let $G$ be a Banach Lie group with Lie algebra $\g = \Lf(G)$, and $H \subseteq G$ a Lie subgroup with Lie algebra $\h = \Lf(H)$. For each $k \in \N_0$, the map \eqref{eq:map_topology_group_expansion} is an analytic isomorphism of Banach Lie groups
	\begin{equation}
		\Theta \colon H \ltimes_\mathrm{Ad} (\g^k \times (\g/\h), \circledast) \rightarrow G^{(k)}_H \, ,
	\end{equation}
	where the second factor in the semidirect product is endowed with the multiplication $\circledast$ defined in \eqref{eq:nustar}, and $G^{(k)}_H$ is the $k$-th Lie group expansion of $G$ with respect to $H$.
\end{thm}

\begin{exmp}
	We are now going to explicitly compute $\circledast$ for the first-order expansion, i.e.\ $k = 1$. From the first terms of the BCH series, we obtain
	\begin{align}
		(\varepsilon c_{(1)} + \varepsilon^2 c_{(2)}) * (\varepsilon d_{(1)} + \varepsilon^2 d_{(2)}) &= (\varepsilon c_{(1)} + \varepsilon^2 c_{(2)}) + (\varepsilon d_{(1)} + \varepsilon^2 d_{(2)}) \nonumber\\
		&\qquad+ \frac{1}{2} [\varepsilon c_{(1)} + \varepsilon^2 c_{(2)}, \varepsilon d_{(1)} + \varepsilon^2 d_{(2)}] + \dots \nonumber\\
		&= \varepsilon (c_{(1)} + d_{(1)}) + \varepsilon^2 \left ( c_{(2)} + d_{(2)} + \frac{1}{2} [c_{(1)}, d_{(1)}] \right ) + \dots
	\end{align}
	Hence, $\circledast$ is given by
	\begin{equation}
		(c_{(1)}, c_{(2)} + \h) \circledast (d_{(1)}, d_{(2)} + \h) = (c_{(1)} + d_{(1)}, c_{(2)} + d_{(2)} + \tfrac{1}{2} [c_{(1)}, d_{(1)}] + \h) \, .
	\end{equation}
\end{exmp}

\enter

Now, we apply this to the example discussed at the beginning of the introduction (\Cref{sec:introduction}):

\begin{exmp} \label{exmp:SO3toISO2}
	As a concrete example, we consider the contraction of $G = \mathrm{SO}(3)$ with respect to the subgroup $H = \{\text{rotations around the $x^3$ axis}\} = \left \{ \begin{psmallmatrix} \cos\alpha& -\sin\alpha& 0\\ \sin\alpha& \cos\alpha& 0\\ 0&0&1 \end{psmallmatrix} \in \mathrm{SO}(3) \right \} \cong \mathrm{SO}(2)$. The corresponding Lie algebras are $\g = \mathfrak{so}(3) = \mathrm{span}\{X_1, X_2, X_3\}$ with $[X_a, X_b] = \tensor{\epsilon}{_{ab}^c} X_c$ and $\h = \mathrm{span}\{X_3\}$.

	Identifying $\g \cong \R^3$ via $x^i X_i \mapsto (x^1, x^2, x^3)$, the Lie bracket on $\g$ corresponds to the vector product on $\R^3$, and the adjoint representation of $G$ on $\g$ corresponds to the defining representation of $\mathrm{SO}(3)$ on $\R^3$. The projection $\g \to \g/\h$ corresponds to the projection $\R^3 \ni (x^1, x^2, x^3) \mapsto (x^1, x^2) \in \R^2$ onto the $x^1$-$x^2$ plane.

	Therefore, the Lie group expansion to order $0$---i.e.\ the Lie group contraction---of $G$ with respect to $H$ is
	\begin{equation}
		G^{(0)}_H \cong H \ltimes_\mathrm{Ad} (\g / \h) \cong \mathrm{SO}(2) \ltimes \R^2 = \mathrm{ISO}(2) \, ,
	\end{equation}
	the Euclidean group in 2 dimensions. Further, the Lie group expansion to order $1$ is
	\begin{subequations}
	\begin{equation}
		G^{(1)}_H \cong \mathrm{SO}(2) \ltimes (\R^3 \times \R^2, \circledast) \, ,
	\end{equation}
	where $\mathrm{SO}(2)$ acts via rotations around the $x^3$ axis on the $\R^3$ factor and naturally on the $\R^2$ factor, and $\circledast$ is (according to the previous example) given by
	\begin{equation}
		(\vec v, \vec w) \circledast (\hat{\vec v}, \hat{\vec w}) = (\vec v + \hat{\vec v}, \vec w + \hat{\vec w} + \tfrac{1}{2} \pi_{\R^2}(\vec v \times \hat{\vec v})) \, .
	\end{equation}
	\end{subequations}
\end{exmp}

\section{Conclusion and outlook} \label{sec:conclusion}

We reviewed general Lie algebra contractions and the particular case of the \.{I}nönü--Wigner procedure. Then, we integrated the latter setup to the corresponding Lie group level using infinite-dimensional Lie theory: Specifically, we reformulated the corresponding setup using Lie algebras of analytic germs in \Cref{prop:Lie_alg_expansion_0_contr} and \Cref{rem:higher-expansions-equivalence}. Next, using their integration theory, we obtained our main results: First, we lifted the construction of Lie algebra expansions via germs to the Lie group level in \Cref{thm:main_result}, and then explicitly identified the expansion groups as Banach Lie groups in \Cref{thm:k-th-expansion}. We exemplified our findings with the following standard example: In the beginning of \Cref{sec:introduction}, we discussed the Lie algebra contraction \(\mathfrak{so}(3) \leadsto \mathfrak{iso}(2)\). Then, in \Cref{exmp:SO3toISO2}, we explicitly calculated the respective first-order Lie group expansion of \(\operatorname{SO}(3) \leadsto \operatorname{ISO}(2)\).

\enter

In a direct follow-up article, we plan to establish an additional perspective on Lie algebra and Lie group contractions using \emph{super} and \emph{graded} Lie theory: Explicitly, Lie algebra expansions of order \(k = 0\) (i.e.\ contractions) inherit the structure of \emph{super} Lie algebras, while expansions truncated at finite \(k \in \N\) inherit the structure of \emph{nilpotent graded} Lie algebras. Notably, super and nilpotent graded Lie algebras can be integrated using so-called \emph{Harish-Chandra modules}, cf.\ \cite{SCHEUNERT2005324,Gavarini}: This corresponds essentially to the semidirect product structure we found in \Cref{thm:k-th-expansion}.

Furthermore, we plan to obtain a \emph{derived} version using the dual perspective of the Chevalley--Eilenberg cochain complex, cf.\ \cite{Jubin_Kotov_Poncin_Salnikov}. Moreover, we want to obtain a similar construction for Lie algebroids and Lie groupoids, as well as \(L_\infty\)-algebras and \(L_\infty\)-groups, cf.\ \cite{Getzler}.

Finally, we want to obtain the Lie group expansions for a \emph{general} Lie algebra contraction map, generalizing the \.{I}nönü--Wigner special case treated here.

\section*{Acknowledgements}

The authors thank the anonymous referee for their careful proofreading, detailed review and valuable feedback. PKS thanks the \emph{Max Planck Institute for Mathematics} for hospitality during his visit in April 2024, where the present project was initiated.

\printbibliography[heading=bibintoc]

\end{document}